\documentclass[twocolumn,prl,floatfix,superscriptaddress,nofootinbib]{revtex4-1}

\usepackage{graphicx}
\usepackage{amsmath,amssymb,bm}
\usepackage{color}
\usepackage{braket}

\graphicspath{{./Figures/}}

\begin{document}

\title{Magnetic resonance detection of individual proton spins using \\ a quantum reporter network}

\author{A. O. Sushkov}
\affiliation{Department of Physics, Harvard University, Cambridge, MA 02138, USA}
\affiliation{Department of Chemistry and Chemical Biology, Harvard University, Cambridge, MA 02138, USA}
\author{I. Lovchinsky}
\affiliation{Department of Physics, Harvard University, Cambridge, MA 02138, USA}
\author{N. Chisholm}
\affiliation{School of Engineering and Applied Sciences, Harvard University, Cambridge, MA 02138, USA}
\author{R. L. Walsworth}
\affiliation{Department of Physics, Harvard University, Cambridge, MA 02138, USA}
\affiliation{Harvard-Smithsonian Center for Astrophysics, Cambridge, Massachusetts 02138, USA}
\affiliation{Center for Brain Science, Harvard University, Cambridge, Massachusetts 02138, USA}
\author{H. Park}
\email{Hongkun\textunderscore Park@harvard.edu}
\affiliation{Department of Physics, Harvard University, Cambridge, MA 02138, USA}
\affiliation{Department of Chemistry and Chemical Biology, Harvard University, Cambridge, MA 02138, USA}
\affiliation{Broad Institute of MIT and Harvard, 7 Cambridge Center, Cambridge, MA, 02142, USA}
\author{M. D. Lukin}
\email{lukin@physics.harvard.edu}
\affiliation{Department of Physics, Harvard University, Cambridge, MA 02138, USA}

\begin{abstract}
We demonstrate a method of magnetic resonance imaging with single nuclear-spin sensitivity under ambient conditions. It employs a network of isolated electronic-spin quantum bits (qubits) that act as quantum reporters on the surface of high purity diamond.  The reporter spins are localized with nanometer-scale uncertainty, and their quantum state is coherently manipulated and measured optically via a proximal nitrogen-vacancy (NV) color center located a few nanometers below the diamond surface. The quantum reporter network is then used for sensing, coherent coupling  and imaging individual proton spins on the diamond surface with angstrom resolution. This approach may enable direct structural imaging of complex molecules that cannot be accessed from bulk studies. It realizes a new platform for probing novel materials, monitoring chemical reactions, and manipulation of complex systems on surfaces at a quantum level.
\end{abstract}

\maketitle

Nuclear magnetic resonance (NMR) and magnetic resonance imaging (MRI) are essential tools for both the physical and life sciences~\cite{Jarenwattananon2013,Logothetis2008}, but have been limited to the detection of large ensembles of spins due to their low sensitivity\cite{Ernst1987,Sakellariou2007}, or the macroscopic nature of sensors~\cite{Nowack2013,Allred2002a}. Over the past decades, significant efforts~\cite{Rugar2004,Degen2009,Taylor2008,Staudacher2013,Mamin2013} have been directed toward pushing this sensitivity to its ultimate physical limit, the detection of individual nuclear spin signals localized in a small volume.
Our approach to magnetic sensing and imaging makes use of a network of electronic spin-1/2 qubits on the surface of a high purity diamond crystal~\cite{Schaffry2011}.
Clean (100) diamond surfaces in ambient conditions are known to host stable electron spins with $S=1/2$ and g-factor of 2~\cite{Grotz2011,Grinolds2014}. These spins have been considered to be deleterious because they are thought to cause decoherence of NV spins within a few nanometers of the diamond surface~\cite{Myers2014,Rosskopf2014}. However, with proper quantum control, these surface electron spins can be turned into a useful resource.
They can be coherently manipulated and measured, serving as a network of quantum ``reporters'' that probe the local magnetic environment. The key advantage of such surface reporter spins is their proximity to sensing targets in samples placed on or near the diamond surface, thereby dramatically enhancing sensitivity and allowing for sub-nanometer localization of individual nuclear spins.

As illustrated in Fig. 1a, we use a single shallow NV center to read out the quantum states of nearby surface reporter spins through the NV-reporter magnetic dipole interaction. The NV center is initialized into the $m_s=0$ sublevel using an optical pumping laser pulse at 532~nm, and the final quantum state of the NV center is read out using its spin-state-dependent fluorescence (Fig. 1b). The spin states of the NV center and of the reporter spins are independently manipulated using pulsed magnetic resonance sequences. The $m_s=0\leftrightarrow m_s=-1$ NV spin transition is addressed at the angular frequency $\omega_{nv}=\Delta-\gamma_e B$, and the $m_s=+1/2\leftrightarrow m_s=-1/2$ surface reporter spin transition is driven at frequency $\omega_s=\gamma_e B$. Here $\Delta=2\pi\times 2.87$~GHz is the NV zero-field splitting, $B$ is the magnitude of the static magnetic field applied along the NV axis, and $\gamma_e=2\pi\times 2.8$~MHz/G is the electron gyromagnetic ratio (Figs. 1c and d).

The magnetic dipole coupling between the NV center and the surface spin network is characterized using a generalized spin-echo (double electron-electron resonance or DEER) sequence, shown in Fig.~1b.
The NV center (NV A) spin-echo
decays on time scale $T^{(nv)}_2 \approx 5$~$\mu$s (Fig. 1e); when a $\pi$-pulse flips the surface reporter spin population simultaneously with the NV-center $\pi$-pulse, the NV-reporter magnetic dipole interaction causes
NV spin echo collapse (Fig. 1e, red circles), with a form that depends on the locations of the surface spins around the NV center. Because the magnetic dipole interaction is long-range, the NV center can be coupled to multiple surface reporter spins, with the coupling strengths dependent on their positions on the diamond surface.
When we treat the diamond with a strongly-oxidizing reflux mixture of concentrated nitric, sulfuric, and perchloric acids~\cite{som} and repeat the DEER experiment on the same NV center, the DEER signal is clearly modified (Fig. 1e, blue triangles), confirming that the reporter spins indeed reside on the diamond surface.
\begin{figure}[h]
 \includegraphics[width=\columnwidth]{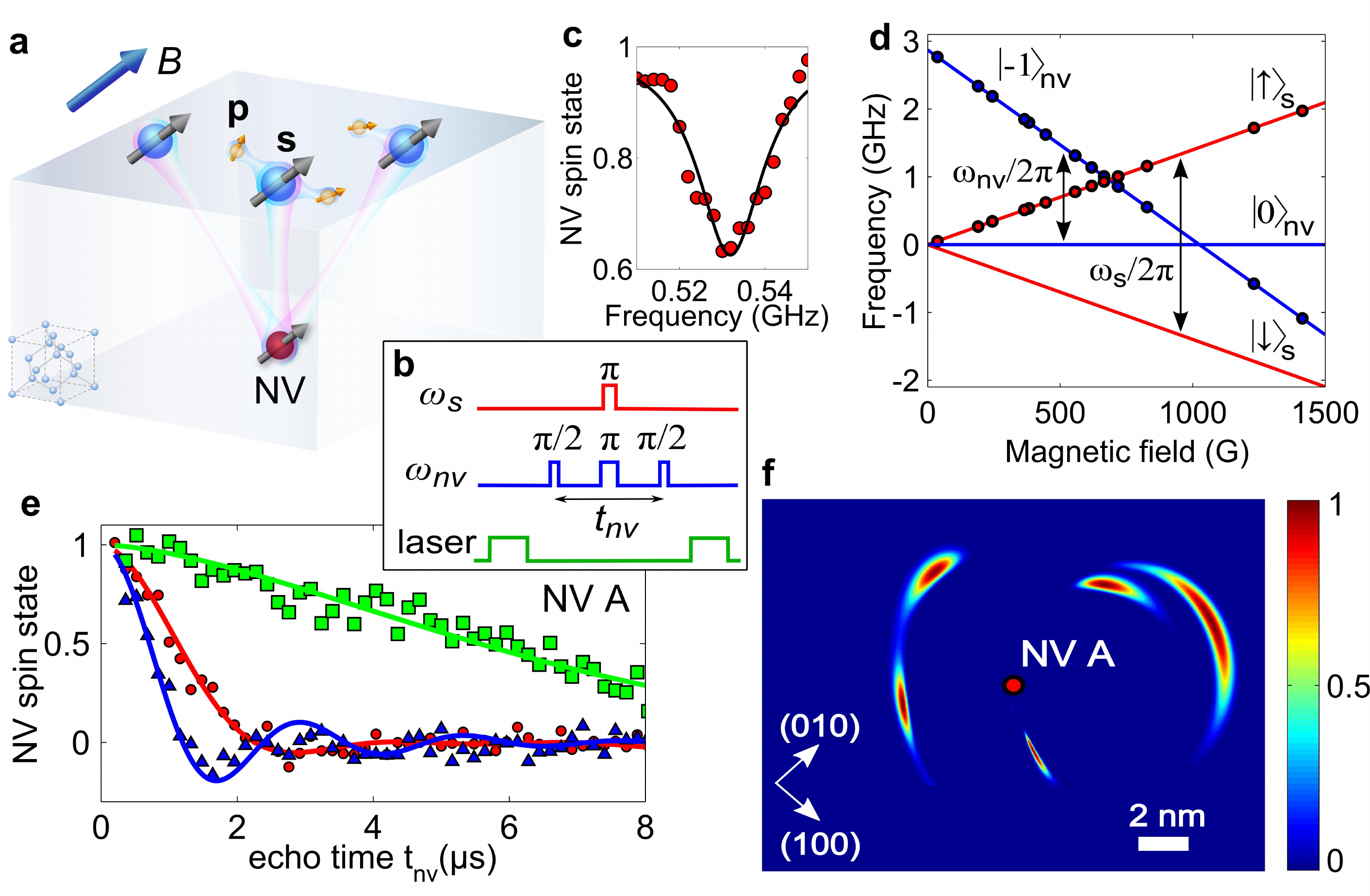}
 \caption{
Characterization of the surface quantum reporter spin network using a shallow NV center.
\textbf{a}, Schematic of a network of reporter electron spins, \textbf{s}, on the surface of a diamond crystal, that can be used to detect and localize surface proton spins, \textbf{p}.
\textbf{b}, DEER pulse sequence.
\textbf{c}, Measured NV DEER signal as a function of reporter spin frequency ($\omega_s/2\pi$) for fixed $t_{nv}$.
\textbf{d}, Measured and calculated Zeeman shifts of NV (blue) and reporter (red) spin states.
\textbf{e}, Results of DEER experiment with varying echo delay time $t_{nv}$. Green squares and line: NV center spin echo decay data and fit. Red circles: DEER measurements. Blue triangles: DEER measurements after oxidizing acid treatment. Red and blue lines are fits using a model with positions of reporter spins on the diamond surface as fitting parameters. In this and subsequent figures spin state populations are scaled to range between -1 and +1.
\textbf{f}, Probability density map for surface reporter spins near NV A, marked by red dot.
Arrows mark diamond crystallographic axes; NV center is aligned along (111).
}
 \label{fig:Figure1}
 \end{figure}

To determine the reporter spin positions, we repeat the DEER measurements while changing the direction of the applied magnetic field $\bm{B}$. The magnetic dipole interaction between the NV center and a surface spin depends on their separation and the angle that the vector between them makes with the vector $\bm{B}$. By rotating $\bm{B}$, we change this angle, and thus the strength of this interaction (similar methods have been employed to localize $^{13}$C spins and other NV centers inside the diamond lattice~\cite{Childress2006,Neumann2010a}). By combining the results of the DEER experiments at 7 different magnetic field angles, we reconstruct the positions of the 4 surface reporter spins nearby the NV center, as shown in Fig. 1f. Here the color scale represents the reporter spin position probability density (normalized to unity), corresponding to the best-fit chi-squared statistic, performed with each reporter spin position fixed at the associated map coordinate~\cite{som}. In particular, the surface reporter spin closest to the NV center can be localized with nanometer-level uncertainty.

The DEER pulse sequence is a useful tool for characterizing the surface reporter spin network on the diamond surface, but it is limited by the decoherence time of the shallow NV center, $T^{(nv)}_2$, which is usually on the order of several microseconds. In order to manipulate and probe the reporter spin network on time scales longer than $T^{(nv)}_2$, we implement a new ``reporter pulse sequence'', shown in Fig. 2a, inset. This protocol, inspired by Ramsey interferometry in atomic physics~\cite{Ramsey1950}, consists of two ``probe'' segments, in which the NV center probes the quantum state of the reporter spin network, separated by an ``evolution'' segment, in which this state can be manipulated. In essence, this protocol enables the comparison of the reporter-spin quantum states before and after the evolution segment. Importantly, the duration of the evolution segment is limited by the NV center $T^{(nv)}_1$ time, rather than its $T^{(nv)}_2$, thereby extending the evolution timescale by orders of magnitude~\cite{Laraoui2013}. In the measurements described below, the duration of the probe segments is kept short ($\approx 0.9$~$\mu$s) to ensure that the NV readout signal is dominated by the coupling to the proximal (most strongly coupled) reporter spin~\cite{som}.

 \begin{figure}[h]
 \includegraphics[width=\columnwidth]{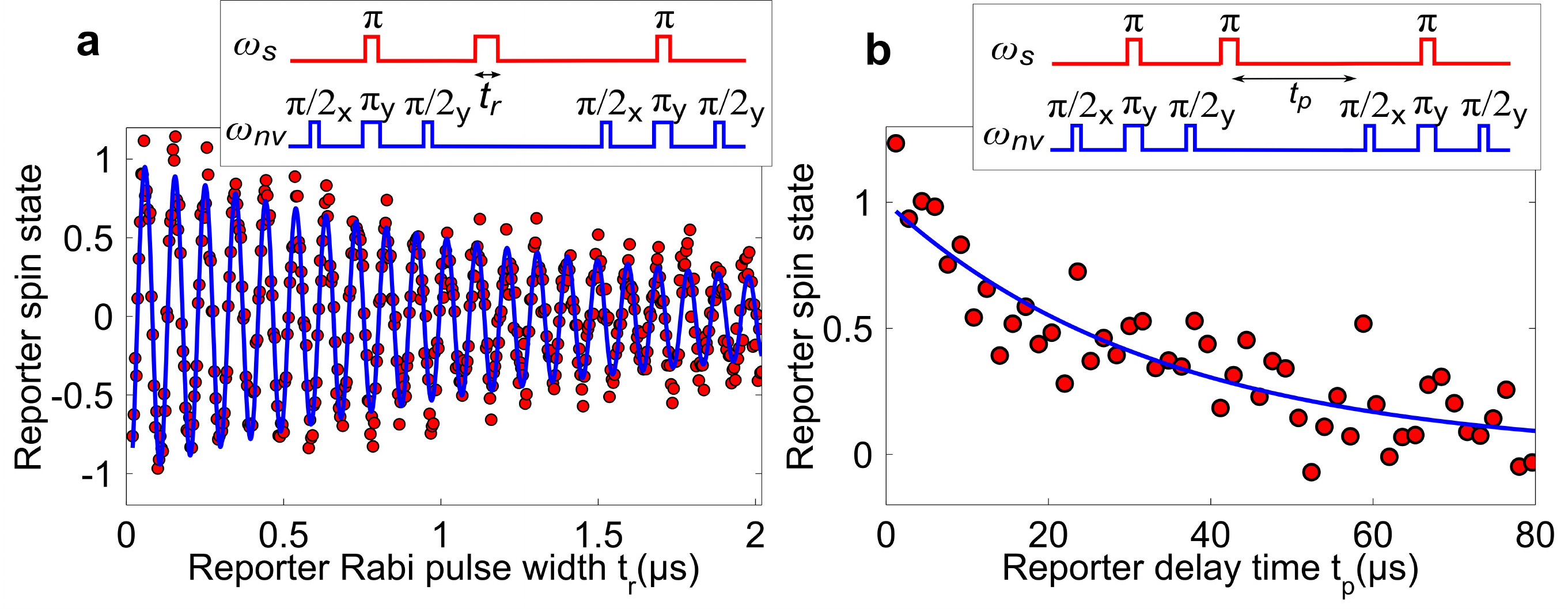}
 \caption{
  \textbf{a}, Coherent control of reporter spins. Rabi oscillations between spin states with a variable-width pulse (red points) with an exponentially-damped fit (blue line). Inset: RF pulse sequence.
  \textbf{b}, Population relaxation dynamics of the reporter spins (red points) with an exponential-decay fit (blue line).
  Inset: RF pulse sequence.
   }
 \label{fig:Figure2}
 \end{figure}

To demonstrate coherent control of the surface reporter spins, we vary the length of the radiofrequency (RF) pulse applied at the reporter spin resonance frequency, as shown in Fig. 2a. We observe Rabi oscillations with decay time on the order of 1~$\mu$s, which is much longer than the reporter-spin Rabi period, indicating that the pulses can be used for coherent control of the reporter spins. Next, the population relaxation time $T^{(s)}_1$ of the surface spin network is measured by varying the delay time $t_p$ between the two probe segments in the pulse sequence, Fig. 2b. The extracted value of $T^{(s)}_1 = (29.4\pm 2.3)$~$\mu$s can be used to place a lower limit of $\approx5$~nm on the mean separation between the surface reporter spins: because, if these spins were closer together, their mutual magnetic dipole flip-flop interaction would give rise to a shorter population relaxation time~\cite{som}. Note that this observation is consistent with the reconstructed spin locations shown in Fig. 1f.

We next use the quantum reporter spin network to perform measurements of the magnetic fields on the diamond surface, using the RF pulse sequence shown in Fig. 3a. The time-varying magnetic field at the site of a reporter spin gives rise to a phase shift during its spin-echo precession time $t_s$, which is converted to a change in its spin state population, and detected by the NV center. By varying the time $t_s$, we implement a frequency filter, whereby the measurement is sensitive to magnetic-field Fourier components at angular frequencies $\omega$ on the order of $2\pi/t_s$, showing up as echo collapses at delay times $t_s=2\pi k/\omega$, where $k=1,3,...$~\cite{Taylor2008}. The experimental data exhibit collapses and revivals characteristic of a time-varying magnetic field created by nuclear spins on the diamond surface, precessing in the applied magnetic field $B$ with Larmor frequency $\omega_n = \gamma_n B$, where $\gamma_n$ is the nuclear spin gyromagnetic ratio. Figure 3a shows example results for a particular NV center (NV A), and the data are consistent with the reporter spin coupled to an oscillating magnetic field created by surface protons with root-mean-squared amplitude of $B_n=0.3$~G and angular frequency of $\omega_n=10.6$~$\mu$s$^{-1}$.
In order to determine the nature of these nuclear spins, we repeat the measurements and analysis at several magnetic fields, and find that the reporter spin echo modulation frequency scales with the applied magnetic field in agreement with the proton gyromagnetic ratio of $2\pi\times 4.26$~kHz/G (Fig. 3b, blue points).

 \begin{figure}[h]
 \includegraphics[width=\columnwidth]{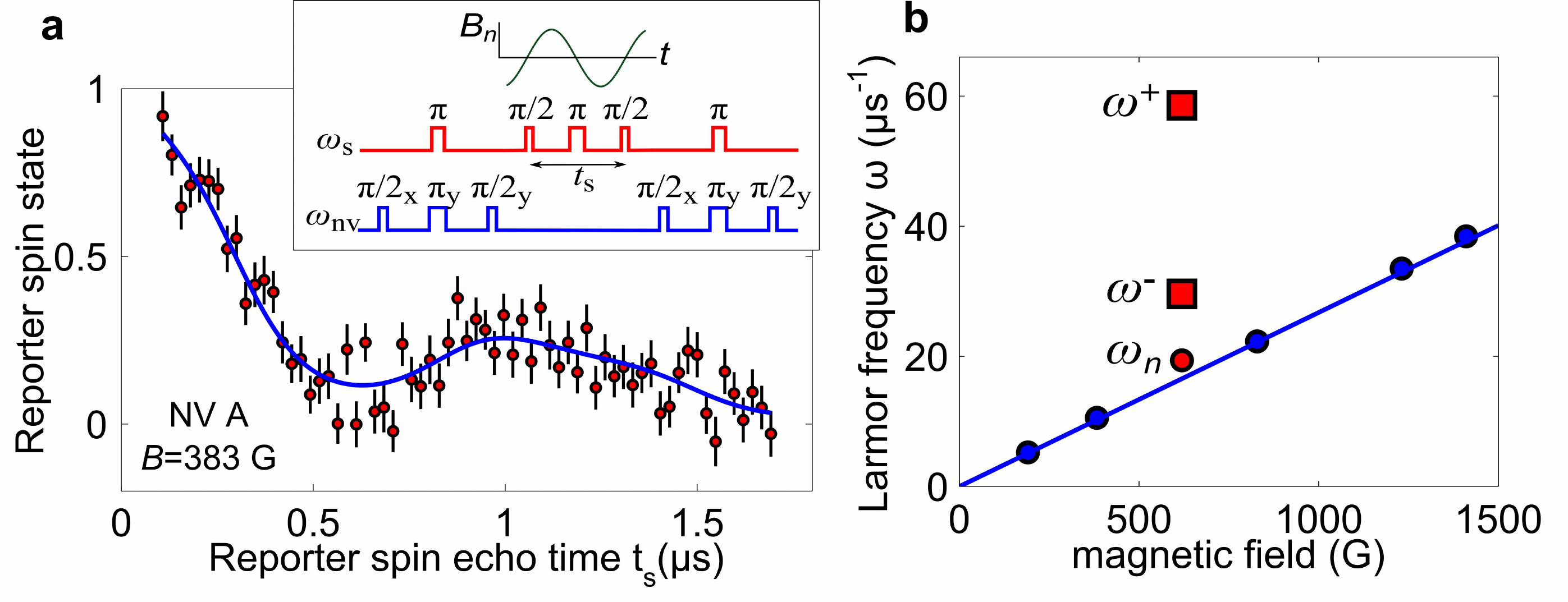}
 \caption{
 Detection of the magnetic field created by protons, using the reporter spins.
  \textbf{a}, Measurement with NV A of the reporter spin echo modulation at $B=383$~G (red points), fit with a model for echo modulation of a reporter spin coupled to a nuclear spin bath~\cite{som} (fit shown by blue line, reduced chi-squared is 1.2).
  The error bars on this and subsequent plots show standard deviations of the data points obtained from averaging approximately 5 million repetitions of the pulse sequence, and are consistent with photon shot noise.
  Inset: reporter echo pulse sequence.
  \textbf{b}, Measured values for $\omega_n$ at 5 different settings of the applied static magnetic field (blue points), consistent with the proton gyromagnetic ratio of $2\pi\times 4.26$~kHz/G (blue line). The red points mark the $\omega_n$, $\omega^-$, and $\omega^+$  oscillation frequencies, see text.
}
 \label{fig:Figure3}
 \end{figure}

Remarkably, however, this simple scaling is not observed at all values of the magnetic field. For example, the data taken with NV A at the magnetic field of 619~G show modulation at frequencies very different from the Larmor frequency expected for the coupling of the reporter spin with a semiclassical proton spin bath (Fig. 3b, red points). This high-frequency modulation, seen in the data plotted in Fig. 4a, signals the presence of coherent dynamics between the reporter and proximal proton spins. In order to reproduce and further explore this coherent coupling, another NV center (NV B) is studied at a similar magnetic field. The experimental points, shown in Fig. 4b, again display strong modulation, crossing the abscissa axis. This signals coherent population transfer between the proton spin states, mediated by the interaction with a single reporter spin, which cannot occur in the absence of reporter/proton entanglement.

To understand these observations, we analyze the coherent dynamics of a reporter electron spin interacting with proximal proton spins on the diamond surface. The hyperfine interaction between them can be described by the Hamiltonian $H = \hbar a J_zI_z + \hbar b J_zI_x$, where $J$ is the spin operator of the reporter qubit, $I$ is the nuclear spin operator, and the $z$-axis is along the applied magnetic field~\cite{Rowan1965,som}.
This Hamiltonian can be interpreted as an effective hyperfine field, created by the reporter spin at the site of the proton spin (Fig. 4c), which in turn gives rise to splitting of the reporter electron spin states,
$\omega^{\pm} = \sqrt{(\pm a/2-\omega_n)^2 + b^2/4}$, as shown in Fig. 4d. This level splitting causes reporter spin echo modulation at frequencies $\omega^+$ and $\omega^-$, with the modulation depth scaling as $2b\omega_n/\omega^+\omega^-$~\cite{Rowan1965}. When the proton Larmor frequency $\omega_n$ is close to half of the hyperfine interaction strength, the reporter spin echo signal is strongly modulated at $\omega^+$ and $\omega^-$, whereas the signal modulation decreases when $\omega_n$ is substantially different from $a,b$.
Data taken at such off-resonance magnetic fields can, within their signal-to-noise ratio, be successfully described with a semiclassical nuclear spin bath model~\cite{som}.

To analyze the experimental data quantitatively, we compare our measurements with a model that includes coherent hyperfine coupling of a reporter electron spin with a proximal proton, as well as the semiclassical spin bath of protons on the diamond surface~\cite{som}. For NV A, the fit to the experimental data shown in Fig.~4a yields spin echo modulation frequencies of $\omega^+=30$~$\mu$s$^{-1}$ and $\omega^-=59$~$\mu$s$^{-1}$, shown as red squares in Fig.~3b, from which we extract hyperfine coupling parameters $a=(66\pm18)$~$\mu$s$^{-1}$ and $b=(52\pm20)$~$\mu$s$^{-1}$.
Both the point magnetic dipole interaction and the contact hyperfine interaction contribute to the parameters $a$ and $b$: $a = a_0 + (\hbar\gamma_e\gamma_n/r_n^3)\left( 1 - 3\cos^2\theta_n \right)$, and $b = (\hbar\gamma_e\gamma_n/r_n^3)\left( 3\cos\theta_n \sin\theta_n \right)$,
where $a_0$ is the contact hyperfine interaction, $r_n$ is the separation between the reporter spin and the proton spin, $\theta_n$ is the angle that the vector between them makes with the applied magnetic field~\cite{som}. The low chemical reactivity of the reporter spins (see discussion below) suggests that the direct overlap between the reporter spin wavefunction and a surface proton is likely minimal, implying that the magnitude of $a_0$ is small.
If we neglect $a_0$, the position of the proximal proton with respect to the reporter spin most strongly coupled to NV A is $r_n=(2.2\pm0.2)$~${\AA}$ and $\theta_n=(26\pm15)^{\circ}$, with the azimuthal angle not quantified by the data.
The data for NV B (Fig. 4b) are consistent with the presence of two proximal protons, coherently coupled to the reporter spin: their best-fit positions are $r_n^{(1)}=(2.6\pm0.2)$~${\AA}$, $\theta_n^{(1)}=(47\pm 3)^{\circ}$ and $r_n^{(2)}=(3.2\pm 0.2)$~${\AA}$, $\theta_n^{(2)}=(19\pm 4)^{\circ}$, as shown in Fig. 4e in a probability density map.
In order to quantify the uncertainty in the proton positions due to our lack of information about the magnitude of the contact interaction, we use $40$~$\mu$s$^{-1}$ as the range of possible values for $a_0$, since this is the measured contact hyperfine interaction with the OH-group proton in a hydroxylated carbon-centered radical~\cite{Rao1965}, similar to our presumed bonding configuration (see below). For this range of $a_0$, the locations of the detected protons are constrained to be within the contours shown in Fig.~4e. We note that in diamond $a_0$ is likely be much smaller due to positive electron affinity of the oxidized diamond surface.

 \begin{figure}[h]
 \includegraphics[width=\columnwidth]{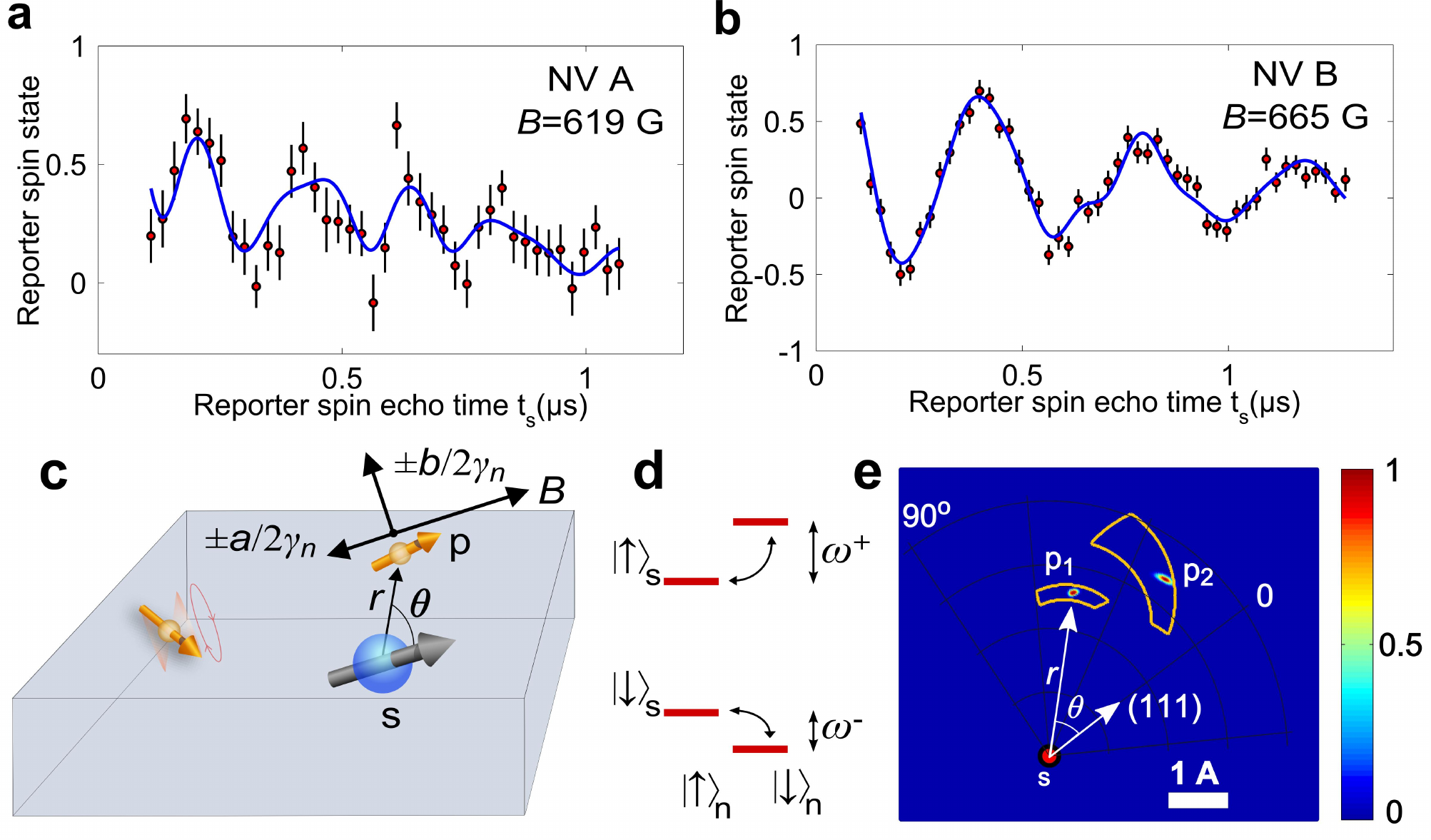}
 \caption{
Coherent dynamics between individual reporter and proton spins.
  \textbf{a},  Reporter spin echo modulation for NV A at $B=619$~G (red points), and fit using a model with the reporter qubit, proximal to the NV center, coupled to one proton spin (blue line).
  \textbf{b}, Reporter spin echo modulation  with NV B at $B=665$~G (red points). The best-fit (blue line, reduced chi-squared value of 1.1) corresponds to a model with the reporter qubit, proximal to the NV center, coupled to two proton spins.
  \textbf{c}, Schematic illustrating hyperfine coupling between the reporter electron spin, s, and the proton spins (gold arrows). The weakly-coupled protons far from the reporter spin precess in the applied magnetic field $B$ at the Larmor frequency. The proximal proton spin, p, experiences the vector sum of $B$ and the effective hyperfine fields $\pm a/2\gamma_n$ and $\pm b/2\gamma_n$, whose signs depend on the reporter spin state.
  \textbf{d}, Energy level diagram for the coupled system of the reporter spin and proximal proton spin.
  \textbf{e}, Localization of the two proximal proton spins (p$_1$ and p$_2$) relative to the reporter spin, s, that is most strongly coupled to NV B. The color scale shows the probability density for each proton location, extracted from a fit to the data shown in (b)~\cite{som}.
  }
 \label{fig:Figure4}
 \end{figure}

While the origin of the reporter spins cannot be unambiguously determined from this study alone, they are likely unsaturated (or ``dangling'') bonds, localized near the top two carbon atom layers. We observe that they are stable in ambient conditions over time scales of many days, which indicates that they are sterically protected from direct chemical reactions with species outside the diamond lattice. Nevertheless, the reporter spins are close to the surface, so that their position changes when exposed to the strongly-oxidizing 3-acid mixture (this surface treatment also modifies the results of the reporter echo experiments, see~\cite{som}); furthermore they can be removed from the diamond surface by annealing the diamond at 465$^{\circ}$C in an O$_2$ atmosphere\cite{Chu2014}. The detected protons are likely from covalently-bound hydroxyl (OH) and carboxyl (COOH) groups terminating the clean diamond surface under ambient conditions~\cite{Mochalin2012}. Their relative locations are consistent with density-functional-theory calculations of the structure of these groups on an oxidized diamond surface~\cite{Sque2006}.

NV centers in diamond have emerged as a nanoscale magnetic-field sensor with exquisite sensitivity under ambient conditions~\cite{Taylor2008,Maze2008,Balasubramanian2008}, enabling magnetic sensing and imaging of single electron spins~\cite{Sushkov2013,Wang2014,Steinert2013} and nanoscale ensembles of nuclear spins~\cite{Staudacher2013,Mamin2013,DeVience2014,Muller2014}.
Our method extends these recent advances into a new domain, enabling magnetic resonance detection and imaging on surfaces with single nuclear spin resolution.
Several paths towards further improving the sensitivity and the broad applicability of our approach should be noted.
It may be possible to extend the reporter spin coherence times using decoupling pulse sequences, together with dilution of the proton magnetic moments on the diamond surface, e.g. by deuteration. Individual addressing of the reporter spins may be achievable with a careful choice of the duration of the ``reporter pulse sequence'' readout intervals, as described above, or via frequency separation of different reporter qubits using a practical magnetic field gradient (less than 1~G/nm)~\cite{Grinolds2014}. Polarization transfer using, for example, Hartmann-Hahn schemes~\cite{Belthangady2013}, from the NV center to the reporter spins, and possibly to strongly coupled surface nuclear spins, may allow initialization and entanglement of the surface spin network~\cite{Goldstein2011} and hyperpolarization of target nuclei. The hyperfine field gradient, produced by the reporter spins, may also be used to encode spatial information for magnetic imaging. Finally, other reporter spin candidates, such as stable nitroxide radicals, can be explored, possibly providing a more flexible route for sensing applications because they can be directly attached to a reactive site of interest on a molecule under study.

Our approach, with  improvements in the coherence properties and robust control of the reporter spins, can enable a number of unique applications. NMR and MRI of individual molecules and proteins under ambient conditions is one direction that can be pursued.
Quantum reporter-based sensing may also find applications in measurements of magnetic fields near complex materials, such as superconductors and topological insulators. Beyond applications in sensing and imaging, our approach provides a powerful new platform for coherent manipulation of coupled electronic and nuclear spins on surfaces or in 2D materials, which can be used to realize and explore  new classes of self-assembled quantum systems~\cite{Cai2013a}.

We thank Eric Kessler, My Linh Pham, Chinmay Belthangady, Nathalie de Leon, Ruffin Evans, and Peter Komar for discussions and experimental help, and Diana Saville for help with graphics.
This work was supported by the Defense Advanced Research Projects Agency (QuASAR program), NSF, CUA, ARO MURI, and Moore Foundation. IL was supported by the AFOSR NDSEG Fellowship, 32 CFR 168a. NC was supported by NSERC PGS D.


%

\end{document}